\documentclass[12pt,thmsa,notitlepage]{article}%
\usepackage{amsmath}
\usepackage{amssymb}
\usepackage{amsfonts}
\usepackage{sw20lart}
\usepackage{setspace}
\usepackage{graphicx}%
\begin{document}

\title{{\LARGE On Vector Goldstone Boson}}
\author{{\large Ling-Fong Li}\\{\small Department of Physics, Carnegie Mellon University, Pittsburgh, PA
15213}}
\maketitle

\begin{abstract}
The possibility that higher dimensional field theories are broken
spontaneously, through the usual Nambu-Goldstone mechanism, to 4-dimension is
explored. As a consequence, vector Goldstone bosons can arise in this breaking
of Lorentzian symmetry from higher dimension to 4-dimension. This can provide
a simple mechanism for reduction to 4-dimension in theories with extra dimensions.

\end{abstract}

\qquad In the 4-dimensional field theory, the spontaneous symmetry breaking is
usually generated by the vacuum expectation values (VEV) of elementary scalar
fields or fermion bilinears in the scalar combination so that the Lorentz
invariance is preserved. Then the fields, which are partners of these fields
under the broken symmetry generators will give rise to massless Goldstone
bosons\cite{Goldstone} \cite{Nambu}. In the case of internal symmetries, the
Goldstone bosons are either scalars or pseduscalars while in the case of
supersymmetries, the Goldstone bosons are fermions, the Goldstinos, which are
supersymmetric partners of the auxiliary scalar fields, $F-$term or $D-$term,
which develop VEV\cite{SUSY}. But there are no vector Goldstone bosons in
these cases because they are not partners of scalar fields under the internal
symmetry operations or space-time symmetries in 4-dimensions. However, in the
theories with extra dimensions\cite{Extra}, the vector fields of four
dimension can be partners of scalar fields, under the Lorentz transformation
in higher dimensional theory. This gives rise to the possibility of vector
Goldstone bosons if we break the Lorentz symmetry of higher dimensional theory
spontaneously to that of 4-dimension\cite{Low}. This can also provide a simple
mechanism for the reduction of higher dimensional theory to the physical
4-dimensional theory. In this paper we will explore this possibility and study
the properties of these vector Goldstone bosons.

Consider a simple case of 5-dimensional field theory where a vector field,
denoted by $\phi_{A},$ will have 5-components, $A=0,1,$ \ldots\ $4.$ The first
4 components, $\phi_{0},\ldots,\phi_{3}$ transform as a vector under
4-dimensional Lorentz transformations while the last component, $\phi_{4}$ is
a scalar. Suppose that in analogy with the 4-dimensional theory the self
interaction of $\phi_{A}$ is of the form,
\begin{equation}
V\left(  \phi_{A}\right)  =\frac{\mu^{2}}{2}\left(  \phi_{A}\phi^{A}\right)
+\frac{\lambda}{4}\left(  \phi_{A}\phi^{A}\right)  ^{2}%
\end{equation}
where $\mu^{2}$ and $\lambda$ are some parameters. The minimal of this
potential is determined by the conditions,
\begin{equation}
\frac{\partial V}{\partial\phi_{A}}=\left[  \mu^{2}+\lambda\left(  \phi
_{B}\phi^{B}\right)  \right]  \phi_{A}=0,\qquad A=0,1,\ldots4
\end{equation}
Thus if any one component of $\phi_{A}$ is non-zero we will have,%
\begin{equation}
\left[  \mu^{2}+\lambda\left(  \phi_{B}\phi^{B}\right)  \right]  =0
\end{equation}
For the case, $\mu^{2}>0,$ $\phi^{2}\equiv\phi_{B}\phi^{B}$ is space-like and
we can choose,%
\begin{equation}
\phi_{4}=v\equiv\sqrt{\frac{\mu^{2}}{\lambda}}\qquad\text{and \ }\phi_{\mu
}=0,\text{ }\mu=0,\ldots3 \label{vev}%
\end{equation}
This breaks the Lorentz symmetry of 5-dimensional theory, $SO\left(
4,1\right)  $ to that of 4-dimensional theory, $SO\left(  3,1\right)  .$ To
find the Goldstone bosons in this case, we write%
\begin{equation}
\phi_{A}^{\prime}=-v_{A}+\phi_{A},\qquad\text{where \qquad}v_{A}=\delta_{A4}v.
\end{equation}
The quadratic terms in the potential is of the form,%
\begin{equation}
V_{2}=\lambda\left(  v\cdot\phi^{\prime}\right)  ^{2}+\frac{1}{2}\phi
^{\prime2}\left[  \lambda\left(  v\cdot v\right)  +\mu^{2}\right]
=\lambda|v|^{2}\phi_{4}^{\prime2}%
\end{equation}
Hence, $\phi_{\mu}^{\prime},$ $\mu=0,\ldots3$ are massless Goldstone bosons
and in this case they transform as vector meson under the 4-dimensional
Lorentz transformations. In other words, these are the vector Goldstone
bosons. It is easy to see that for the case $\mu^{2}<0,$ $\phi^{2}$ is
time-like and the symmetry breaking is from $SO\left(  4,1\right)  $ down to
$SO\left(  4\right)  .$ This is not physically interesting, because the
resulting theory will not have Lorentz symmetry in 4-dimension.

To study the broken symmetry generators, we write down the commutation
relations between the Lorentz generators $W^{AB}$ and the vector fields
$\phi_{C}$,%
\begin{equation}
\left[  W_{AB},\phi_{C}\right]  =ig_{AC}\,\phi_{B}-ig_{BC}\,\phi_{A},\qquad
A,B,C=0,1,\ldots4
\end{equation}
where $g_{AB}=\left(  1,-1,-1,-1,-1\right)  $ is the metric for the
5-dimensional space. In particular, we have
\begin{equation}
\left[  W_{\mu4},\phi_{\nu}\right]  =-i\,g_{\mu\nu}\phi_{4},\qquad\mu
,\nu=0,1,2,3
\end{equation}
Then from the usual Goldstone theorem \cite{Goldstone}, $\phi_{\mu},$
$\mu=0,1,2,3$ are massless and $W_{\mu4},\mu=0,1,2,3$ are the broken
generators. It is not hard to see that these vector Goldstone bosons
$\phi_{\mu}$ coupled to the densities of the Lorentz generators, $M_{\alpha
\mu4}.$ This is analogous to the case of spontaneous breaking of chiral
symmetry where Goldstone pions , $\pi^{\alpha}$ ,couple to the axial vector
currents $A_{\mu}^{\alpha},$ which are the densities of the broken generators.

If the vector Goldstone boson can couple to the fermion, the interaction will
be of the form%
\begin{equation}
L_{V\psi}=f\overset{\_}{\psi}\gamma_{A}\psi\phi^{A}%
\end{equation}
where the 5-dimensional gamma matrices are of the form%
\begin{equation}
\gamma^{A}=\left(  \gamma^{0},\gamma^{1},\gamma^{2},\gamma^{3},i\gamma
^{5}\right)
\end{equation}
Then the spontaneous symmetry breaking, in Eq(\ref{vev}), will contribute to
the fermion masses. Note that the $\gamma_{5}$ factor in the fermion bilinear
can be removed by a chiral rotation on the fermion field. On interesting
feature here is that the Goldstone mode will have coupling to some combination
of vector and axial vector currents. Physical consequences of this type of
coupling might be of interest phenomenologically. Similarly, a coupling of
$\phi_{A}$ to a scalar field of the form,%
\[
L_{V\phi1}=f^{\prime}\phi_{A}\phi^{A}\phi\phi
\]
can give contribution to scalar mass and coupling of the Goldstone boson of
the form,%
\[
\phi_{\mu}\phi^{\mu}\phi\phi.
\]
Another possible type of coupling is the derivative coupling,%
\[
L_{V\phi2}=g\phi\phi_{A}\partial^{A}\phi
\]
which can give a coupling of the form,%
\[
\phi\phi_{\mu}\partial^{\mu}\phi.
\]
in 4-dimension.

The generalization to six or higher dimensional theory is straightforward. But
the structure of the symmetry breaking will be more complicate. For example,
to reduce 6-dimensional Lorentzian symmetry, $SO\left(  5,1\right)  $ to
$SO\left(  3,1\right)  $ \ of the 4-dimensional Lorentz symmetry, we can use
two 6-dimensional vector fields in analogy with the breaking of the internal
symmetries. It is also possible to break the higher dimensional Lorentz
symmetry by using higher rank tensor fields\cite{LFLI}. It is conceivable that
higher dimensional Lorentzian symmetry can be broken down to $SO\left(
3,1\right)  \times G$ where $G$ is some compact internal symmetry group.

We can also explore the cases where these vector fields in higher dimension
also carry some internal quantum numbers. Again consider the simple case of
5-dimensional theory. Let $\phi_{i}^{A},i=1,2,\ldots n$ be a set of vector
fields which transform as fundamental representation under the internal
symmetry group $SO\left(  n\right)  .$The effective potential is then of the
form\cite{LFLI}%
\begin{equation}
V=\frac{\mu^{2}}{2}\left(  \phi_{i}^{A}\phi_{Ai}\right)  +\frac{\lambda_{1}%
}{4}\left(  \phi_{i}^{A}\phi_{Ai}\right)  ^{2}+\frac{\lambda_{2}}{4}\left(
\phi_{i}^{A}\phi_{Aj}\right)  \left(  \phi_{i}^{B}\phi_{Bj}\right)
\end{equation}
which has the symmetry, $SO\left(  4,1\right)  \times SO\left(  n\right)  .$
Using the results from the breaking of the internal symmetry\cite{LFLI}, we
can deduce that for the case $\lambda_{2}<0$ the symmetry breaking \ has the
pattern,%
\[
SO\left(  4,1\right)  \times SO\left(  n\right)  \rightarrow SO\left(
3,1\right)  \times SO\left(  n-1\right)
\]
Presumably, there will be 1 vector Goldstone boson for the broken generators,
$W_{\mu4},$ and $n-1$ scalar Goldstone bosons for the broken internal symmetry
generators. Thus in this simple case there is no connection between Lorentzian
and internal symmetries. It is conceivable that in more complicate cases there
might be some coupling between internal and Lorentzian symmetry. Recall that
in the usual spontaneous symmetry breaking of the chiral $SU\left(  3\right)
_{L}\times SU\left(  3\right)  _{R}$ symmetry in the low energy hadronic
interaction down to $SU\left(  3\right)  _{V},$\cite{Chiral} the broken
generators are the axial charges which is the linear combination of left and
right generators. If we replace replace one of the the internal symmetry by
\ Lorentzian symmetry, it is possible that some combination of Lorentz
symmetry and internal symmetry generators are broken. In this case, the vector
Goldstone bosons can carry the internal symmetry quantum numbers.

So far we have used elementary fields to breaking the symmetry spontaneously.
It is clear that similar breaking can be generated by the composite fields.
For example, in 5-dimension condensation of the fermion bilinears of the form,%
\begin{equation}
\left\langle \overset{\_}{\psi}\gamma_{A}\psi\right\rangle =v\delta_{A4}%
\end{equation}
can also breaking the Lorentzian symmetry and gives vector Goldstone boson.

This work is supported in part by U.S. Department of Energy (Grant No. DE-FG
02-91 ER 40682).

\end{document}